\documentclass[aps,prd,preprint,superscriptaddress,tightenlines,nofootinbib]{revtex4}

\usepackage{graphicx}
\usepackage{dcolumn}
\usepackage{bm}

\graphicspath{ {.}
               {/cdat/lnspu4/disk1/huanggs/run/PsippProc/}
             }

\newcommand{\FourPi}   {$2(\pi^+ \pi^-)$}
\newcommand{\RhoPiPi}  {$\rho \pi^+ \pi^-$}

\newcommand{\FivePi}   {$2(\pi^+ \pi^-)\pi^0$}
\newcommand{\EtaPiPi}  {$\eta \pi^+ \pi^-$}
\newcommand{\OmegaPiPi}{$\omega \pi^+ \pi^-$}

\newcommand{\EtaTriPi} {$\eta 3\pi$}
\newcommand{\EtaTriPiA}{$\eta 3\pi(\eta\to\gamma\gamma)$}
\newcommand{\EtaTriPiB}{$\eta 3\pi(\eta\to 3\pi)$}
\newcommand{\EtapTriPi}{$\eta^\prime 3\pi$}
\newcommand{\KKPiPi}   {$K^+K^-\pi^+\pi^-$}
\newcommand{\RhoKK}    {$\rho K^+K^-$}
\newcommand{\PhiPiPi}  {$\phi \pi^+ \pi^-$}

\newcommand{\KKTriPi}  {$K^+K^-\pi^+\pi^-\pi^0$}
\newcommand{\EtaKK}    {$\eta K^+ K^-$}
\newcommand{\OmegaKK}  {$\omega K^+ K^-$}
\newcommand{\KKKK}     {$2(K^+K^-)$}
\newcommand{\PhiKK}    {$\phi K^+ K^-$}
\newcommand{\KKKKPi}   {$2(K^+K^-)\pi^0$}
\newcommand{\PPPiPi}   {$p \bar{p} \pi^+ \pi^-$}
\newcommand{\RhoPP}    {$\rho p \bar{p}$}
\newcommand{\PPTriPi}  {$p \bar{p} \pi^+ \pi^-\pi^0$}
\newcommand{\EtaPP}    {$\eta p \bar{p}$}
\newcommand{\OmegaPP}  {$\omega p \bar{p}$}
\newcommand{\PPKK}     {$p \bar{p} K^+ K^-$}
\newcommand{\PhiPP}    {$\phi p \bar{p}$}

\newcommand{\LLPiPi}   {$\Lambda\bar\Lambda\pi^+\pi^-$}
\newcommand{\LamPK}    {$\Lambda \bar{p} K^+$}
\newcommand{\LamPKPiPi}{$\Lambda\bar{p}K^+\pi^+\pi^-$}
\def\Journal#1&#2&#3(#4){#1{\bf #2}, #3 (#4)}
\def\NIMA{Nucl. Instrum. Methods Phys. Res., Sect. A }
\def\NPB{Nucl.  Phys.  B }
\def\PLB{Phys.  Lett.  B }
\def\PRL{Phys.  Rev.  Lett.  }
\def\PRD{Phys.  Rev.  D }

\def\etal{{\it et al.}}

\begin{document}
\preprint{CLNS 05/1917}       
\preprint{CLEO 05-9}         

\title{\bf Observation of Thirteen New Exclusive Multi-body Hadronic Decays of the $\psi(2S)$}

\author{R.~A.~Briere}
\author{G.~P.~Chen}
\author{J.~Chen}
\author{T.~Ferguson}
\author{G.~Tatishvili}
\author{H.~Vogel}
\author{M.~E.~Watkins}
\affiliation{Carnegie Mellon University, Pittsburgh, Pennsylvania 15213}
\author{J.~L.~Rosner}
\affiliation{Enrico Fermi Institute, University of
Chicago, Chicago, Illinois 60637}
% ...
\collaboration{CLEO Collaboration} 
\noaffiliation

\date{August 3, 2005}

\begin{abstract}
Using data accumulated with the CLEO detector corresponding to an
integrated luminosity of $\cal{L}$=5.63~pb$^{-1}$ on the peak of
the $\psi(2S)$ ($3.08\times 10^6$ $\psi(2S)$ decays)
and 20.70~pb$^{-1}$ at $\sqrt{s}$=3.67~GeV, we
report first measurements of the branching fractions for the
following 13 decay modes of the $\psi(2S)$: \EtaTriPi,
\EtapTriPi, \RhoKK, \KKTriPi, \KKKK, \KKKKPi, \RhoPP, \PPTriPi, \EtaPP,
\PPKK, \LLPiPi, \LamPK, and \LamPKPiPi, and more precise
measurements of 8 previously measured modes: \FourPi,
\RhoPiPi, \FivePi, \OmegaPiPi, \KKPiPi, \OmegaKK, \PhiKK, and
\PPPiPi. We also report new branching fraction measurements of
\PhiPiPi\ and \OmegaPP\ and upper limits for \EtaPiPi, \EtaKK\ and
\PhiPP.
Results are compared, where possible, with the corresponding
$J/\psi$ branching ratios to provide new tests of the 12\% rule.

\end{abstract}

\maketitle

The states $J/\psi$ and $\psi(2S)$ are non-relativistic bound
states of a charm and an anti-charm quark. In perturbative QCD the 
decays of these states are expected to be dominated by the 
annihilation of the constituent $c\bar{c}$ into three gluons
or a virtual photon. 
The partial width for the decays into an exclusive hadronic state $h$
is expected to be proportional to the square of the
$c\bar{c}$ wave function overlap at zero quark separation, 
which is well determined from the leptonic width~\cite{PDG}. 
Since the strong coupling constant, $\alpha_s$, is not very
different at the $J/\psi$ and $\psi(2S)$ masses, it is expected that
for any state $h$ the
 $J/\psi$ and $\psi(2S)$ branching ratios are related by~\cite{RULE}
\begin{equation}
Q_h=\frac{{\cal B}(\psi(2S)\to h)}{{\cal B}(J/\psi\to h)}
\approx
\frac{{\cal B}(\psi(2S)\to \ell^+\ell^-)}{{\cal B}(J/\psi\to\ell^+\ell^-)}
=(12.7 \pm 0.5)\%,
\label{equ:q}
\end{equation}
where ${\cal B}$ denotes a branching fraction,
and the leptonic branching fractions are taken from the 
Particle Data Group (PDG)~\cite{PDG}.
This relation is sometimes called \lq \lq the 12\% rule''. Modest
deviations from the rule are expected~\cite{GULI}. Although the
rule works well for some specific decay modes of the $\psi(2S)$,
isospin conserving $\psi(2S)$ decays to two-body final states 
consisting of one vector and one pseudsoscalar meson exhibit strong
suppression: $\rho\pi$ is suppressed by a factor of seventy compared 
to the expectations of the rule~\cite{PDG,MOREBES,HARRIS,BHHMK}.
Also, vector-tensor channels such as $\rho a_2(1320)$, 
and $K^*(892)\bar{K}^*_2(1430)$ are significantly
suppressed \cite{PDG,BESVT}.
A recent review~\cite{GULI} of relevant
theory and experiment concludes that current theoretical explanations are
unsatisfactory.
Clearly, more experimental results are desirable.

This Letter presents new measurements of a wide selection of
$\psi(2S)$ decays including modes with and without strange
particles and with and without baryons.  The following modes of
the $\psi(2S)$ are observed for the first time: \EtaTriPi,
\EtapTriPi, \RhoKK, \KKTriPi, \KKKK, \KKKKPi, \RhoPP, \PPTriPi, \EtaPP,
\PPKK, \LLPiPi, \LamPK, and  \LamPKPiPi. More precise
branching fraction measurements of previously measured modes 
\FourPi, \RhoPiPi, \FivePi, \OmegaPiPi, \KKPiPi, \OmegaKK, \PhiKK,
and \PPPiPi\ are also reported. We also measure \PhiPiPi\ and
\OmegaPP\ and obtain upper limits for \EtaPiPi, \EtaKK\ and
\PhiPP. Where applicable, the inclusion of charge conjugate modes
is implied. As 14 of the modes we study have been previously
observed at the $J/\psi$, 14 tests of the 12\% rule are
made, 5 for the first time and 9 with greater precision than
corresponding previous tests.

The data sample used in this analysis is obtained at the $\psi(2S)$ and the nearby continuum
in $e^+e^-$ collisions produced by the Cornell Electron Storage Ring (CESR) and acquired with
the CLEO detector.
The CLEO~III detector~\cite{cleoiiidetector}
features a solid angle
coverage for charged and neutral particles of 93\%.
The charged particle tracking system, operating in a
1.0~T magnetic field along the beam axis, achieves
a momentum resolution of $\sim$0.6\% at
$p=1$~GeV/$c$. The calorimeter attains a photon
energy resolution of 2.2\% at $E_\gamma=1$~GeV and 5\% at 100~MeV.
Two particle identification systems, one based on energy loss ($dE/dx$) in
the drift chamber and the other a ring-imaging Cherenkov (RICH)
detector, are used together to separate $K^\pm$ from $\pi^\pm$.
The combined $dE/dx$-RICH particle identification procedure has
efficiencies exceeding 90\% and misidentification rates below 5\%
for both $\pi^\pm$ and $K^\pm$ for the momenta below 2.5 GeV/$c$.

Half of the $\psi(2S)$ data and all
the continuum data were taken after
a transition to the CLEO-c~\cite{YELLOWBOOK} detector configuration, in which
CLEO~III's silicon-strip vertex detector was replaced with a six-layer
all-stereo drift chamber.
The two detector configurations also correspond
to different accelerator lattices: the former with
a single wiggler magnet and a center-of-mass
energy spread of 1.5~MeV, the latter
(CESR-c~\cite{YELLOWBOOK}) with
six wiggler magnets and an energy spread of 2.3~MeV.

The integrated luminosity ($\cal{L}$) of the datasets was measured
using $e^+ e^-$, $\gamma\gamma$, and $\mu^+ \mu^-$ final
states~\cite{LUMINS}. Event counts were normalized with a Monte
Carlo (MC) simulation based on the Babayaga~\cite{BBY} event
generator combined with GEANT-based~\cite{GEANT} detector
modeling. The data consist of  $\cal{L}$=5.63~pb$^{-1}$ on the
peak of the $\psi(2S)$ (2.74~pb$^{-1}$ for CLEO~III,
2.89~pb$^{-1}$ for CLEO-c) and 20.70~pb$^{-1}$ at $\sqrt{s}$=3.67~GeV 
(16 MeV below the $\psi(2S)$, all CLEO-c). The nominal scale factor used to
normalize continuum yields to $\psi(2S)$ data is $f_{\rm
nom}=0.270\pm0.005$, and is determined from the integrated
luminosities of the data sets corrected for the $1/s$ dependence
of the cross section, where the error is from the relative
luminosity uncertainty, and the uncertainty in the $s$ dependence
of the cross section. (The scale factor does not differ
appreciably if a high power of $1/s$ is used: a factor of 0.87\% for each
power). The value of $f$ used for each mode also corrects for the
small difference in efficiency between the $\psi(2S)$ and
continuum data.

Standard requirements are used to select charged particles
reconstructed in the tracking system and photon candidates in the
CsI calorimeter. We require tracks of charged particles
to have momenta $p>100$~MeV/$c$
and to satisfy $|\cos\theta|<0.90$, where $\theta$ is the polar angle
with respect to the $e^+$ direction.
Each photon candidate satisfies  $E_\gamma>30$~MeV and is more than 8\,cm
away from the projections of tracks into the calorimeter.
Particle identification from $dE/dx$ and the RICH detector is used on 
all charged particle candidates. Pions, kaons, and protons must be 
positively and uniquely identified. That is: 
pion candidates must not satisfy kaon or proton selection criteria,
and kaon and proton candidates obey similar requirements.
Charged particles must not be identified as electrons using criteria
based on momentum, calorimeter energy deposition, and $dE/dx$.

The invariant mass of the decay products from the following particles
must lie within limits determined from MC studies:
$\pi^0~(120 \le M_{\gamma\gamma} \le 150 {\rm ~MeV})$, 
$\eta ~(500 \le M_{\gamma\gamma} \le 580 {\rm ~MeV})$, 
$\eta ~(530 \le M_{\pi^+\pi^-\pi^0} \le 565 {\rm ~MeV})$, 
$\omega ~(740 \le M_{\pi^+\pi^-\pi^0} \le 820 {\rm ~MeV}$ 
[$760 \le M_{\pi^+ \pi^- \pi^0} \le 800 {\rm ~MeV}$ 
for the $\omega p \bar p$ final state]), 
$\phi ~(1.00 \le M_{K^+K^-} \le 1.04 {\rm ~GeV})$, and 
$\Lambda ~(1.112 \le M_{p \pi^-} \le 1.120 {\rm ~GeV})$. 
For $\pi^0 \rightarrow \gamma \gamma$
and $\eta \rightarrow \gamma \gamma$ candidates in events with
more than two photons, the combination giving a mass closest to
the known  $\pi^0$ or $\eta$ mass is chosen, and a kinematically
constrained fit to the known parent mass is made. Fake $\pi^0{\rm
's}$ and $\eta{\rm 's}$ are suppressed by requiring that each
electromagnetic shower profile be consistent with that of a
photon.
For $\eta \rightarrow \pi^+ \pi^- \pi^0$ and 
$\omega \rightarrow \pi^+ \pi^- \pi^0$, 
the $\pi^0$ is selected as described above, and then
combined with all possible combinations of two oppositely charged
pions choosing the combination that is closest to the $\eta
(\omega)$ mass. A kinematically constrained fit is not used for
either of these modes, or for $\phi{\rm's}$ or for
$\Lambda{\rm's}$. For $\Lambda \rightarrow p \pi^-$, a fit of the
$p \pi^-$ trajectories to a common vertex separated from the $e^+
e^-$ interaction ellipsoid is made. Contamination from $K^0_S$
decays is eliminated by the energy and momentum requirements
imposed on the event, and by particle identification.

Energy and momentum conservation requirements are imposed on the
reconstructed final state hadrons, which have momentum $p_i$ and
combined measured energy $E_{\rm vis}$. We require the measured
scaled energy $E_{\rm vis}$/$E_{\rm c.m.}$ be consistent with unity
within experimental resolution, which varies by final state.
We require $|\sum {\bf p_i}|/E_{\rm c.m.}< 0.02$. Together these
requirements suppress backgrounds with missing energy or incorrect
mass assignments. The experimental resolutions are smaller than
1\% in scaled energy and 2\% in scaled momentum.

For the final states with four charged tracks and a $\pi^0$,
an additional cut is applied to remove a background of radiative
events. When the photon is combined with a low-energy photon candidate,
it can imitate a $\pi^0$. We require
$(E_{\rm 4 tracks}+E_\gamma)$/$E_{\rm c.m.} < 0.995$,
where $E_\gamma$ is the energy of the highest energy photon.

In order to compute $Q_h$ in modes with two or more charged pions,
two $\pi^0{\rm's}$ or an $\eta$, it is necessary to remove the
contribution from the transitions $\psi(2S) \rightarrow J/ \psi
X$, where $X= \pi^+ \pi^-$, $\pi^0 \pi^0$, or $\eta$~\cite{note}.
Accordingly we reject events in which the mass of any of the
following falls within the range
$3.05<m<3.15$~GeV~\cite{eta_range}: the two highest momentum
oppositely charged tracks, the recoil mass against the two lowest
momentum oppositely charged tracks, or the mass recoiling against
the 2$\pi^0{\rm's}$ or $\eta$.


For every  final state, a signal selection range in 
$E_{\rm vis}/E_{\rm c.m.}$ is determined by Monte Carlo simulation, 
and a sideband selection range is defined to measure background. 
Final states with an intermediate $\rho$,  $\eta$, 
$\eta^\prime$, $\omega$, $\phi$ or $\Lambda$
must satisfy a scaled energy signal selection range requirement
identical to the corresponding mode without the intermediate
particle. For example, the scaled energy signal selection range is
the same for $\phi K^+K^-$ and $K^+K^-K^+K^-$. For final states
with an $\eta$, $\eta^\prime$,  $\omega$, $\phi$, or $\Lambda$
the event yield is determined from signal and sideband selection
ranges of the particle mass. For final states with a $\rho$ the
event yield is determined from a fit to the $\pi^+ \pi^-$ invariant
mass by a Breit-Wigner with parameters taken from the PDG~\cite{PDG}
(for \RhoPiPi, the same-sign $\pi^\pm \pi^\pm$ invariant mass
distribution is used to estimate the background); 
an exception is \RhoPP, where the yield is determined from the 
signal region: $0.54 \le m_{\pi\pi} \le 1.0$ GeV assuming 
a linear background. Most modes studied in this Letter have resonant
sub-modes, however; we only report branching ratios for resonant
sub-modes which can be cleanly separated from background.

\begin{figure}[htbp]
  \centering
  \includegraphics[height=0.35\textheight]
   {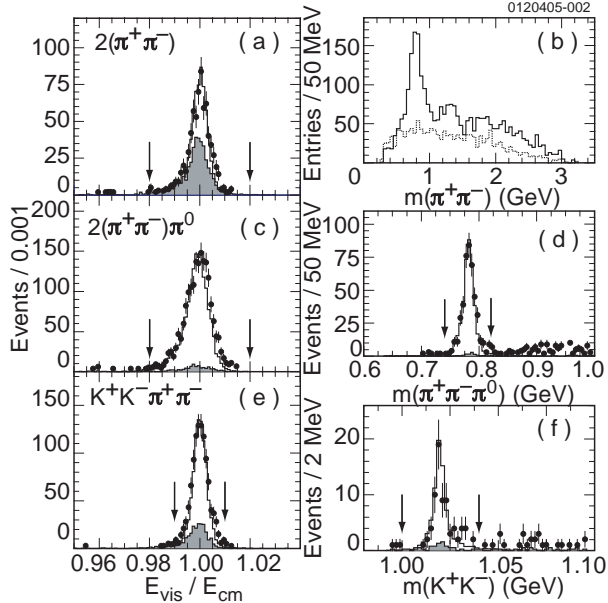}
  \caption{Distributions for the modes \FourPi ~(a) and (b),  \FivePi ~(c) and (d),
  and \KKPiPi ~(e) and (f). The pairs of arrows indicate the signal
  regions.
           (a), (c), and (e) The scaled total energy (filled
           circle with error bar: $\psi(2S)$ data,
               solid line: Monte Carlo, shaded histogram:
               continuum).
           (b) The $\pi^+\pi^-$ invariant mass in \FourPi ~(solid
           line: data, dashed line: like sign $\pi^\pm \pi^\pm$ invariant
           mass, which is used to estimate combinatorial
           background).
           (d) The $\pi^+ \pi^- \pi^0$ invariant mass in \FivePi
           (filled circle with error bar: $\psi(2S)$ data,
               solid line: Monte Carlo, shaded histogram: continuum).
           (f) The $K^+K^-$ invariant mass in \KKPiPi.
               (The symbol key is the same as (d).) }
  \label{fig:selected}
\end{figure}

In Fig.~\ref{fig:selected} the scaled energy and invariant mass
distributions are shown for three
typical modes: \FourPi, \FivePi, and \KKPiPi. Evidence for
production of the $\rho$, $\omega$ and $\phi$ resonances,
respectively, is observed in the corresponding mass spectra.

Event totals are shown for both the $\psi(2S)$ and
the continuum in Table \ref{tab:br}, where $S_{\psi(2S)}$ $(S_{co})$
is the number of events in the signal region and  $B_{\psi(2S)}$ $(B_{co})$
the number of events in the sideband region in $\psi(2S)$ (continuum) data.
Under the assumption that interference between $\psi(2S)$ decay
and continuum production of the same final state
is absent, the number of events attributable to each $\psi(2S)$ decay mode,
$N_S$, is
\begin{equation}
N_S = S_{\psi(2S)} - B_{\psi(2S)} - f(S_{co} - B_{co}),
\label{equ:ns}
\end{equation}
where $f$ is mode dependent and listed in Table \ref{tab:br}.
We observe a statistically significant signal in all modes except
\EtaPiPi, \EtaKK\ and $\phi p \bar p$.
The signal with the least statistical significance is \EtaPP\ $(4.2 \sigma)$.

The efficiency, $\epsilon$, for each final state is the average
obtained from MC simulations~\cite{GEANT} for both detector configurations;
the two values are typically within a few percent (relative) to each other.
No initial state radiation is included in the Monte Carlo, but final state
radiation is accounted for. The efficiencies in Table \ref{tab:br} include
the branching ratios for intermediate final states.

\begin{table*}[!h]
\caption{For each final state $h$ the following quantities are given: 
the number of events in the signal region, $S_{co}$, 
and background from sidebands, $B_{co}$, in continuum data; 
the scale  factor, $f$; the number of events in
the signal region, $S_{\psi(2S)}$, and background from sidebands,
$B_{\psi(2S)}$, in $\psi(2S)$ data; the number of events
attributable to $\psi(2S)$ decay, $N_S$, computed according to
Eq.~\ref{equ:ns}; the average efficiency, $\epsilon$; the absolute
branching fraction with statistical (68\% C.L.) and systematic
errors; previous branching fraction measurements from the
PDG~\cite{PDG}, and the $Q_h$ value.
For $\eta 3 \pi$, the two decays modes $\eta 3 \pi(\eta \to \gamma\gamma)$
and  $\eta 3 \pi(\eta \to 3 \pi)$
are combined on line $\eta 3\pi$.
}
\begin{center}
\begin{tabular}{|c|c|c|c|c|c|c|c|c|c|c|}  \hline
mode
          &  $S_{co}$  & $B_{co}$
                                         & $f$
                                                 & $S_{\psi(2S)}$ & $B_{\psi(2S)}$
                                                              & $N_S$
                                                                       &$\varepsilon$
&          ${\cal B}(\psi(2S) \to h)$  & ${\cal B}$ (PDG) & $Q_h$ \\
$h$   &  & & & & & &  &  (units of $10^{-4}$) & (units of $10^{-4}$)    &(\%) \\
\hline
\FourPi   &1471 & 28 & 0.2668 &  713 & 20 & 308.0& 0.4507 &  2.2$\pm$0.2$\pm$0.2 &  4.50$\pm$1.00 &  5.55$\pm$1.53 \\
\RhoPiPi  &1168 &  - & 0.2667 &  597 &  - & 285.5& 0.4679 &  2.0$\pm$0.2$\pm$0.4 &  4.20$\pm$1.50 &        -       \\
\FivePi   & 352 & 25 & 0.2550 & 1825 & 39 &1702.6& 0.2115 & 26.1$\pm$0.7$\pm$3.0 & 30.00$\pm$8.00 &  7.76$\pm$1.10 \\
\EtaPiPi  &  15 &  0 & 0.2501 &   13 &  2 &   7.2& 0.0416 &     $ < 1.6 $        &        -       &        -       \\
\OmegaPiPi&  43 &  9 & 0.2357 &  437 & 38 & 391.0& 0.1553 &  8.2$\pm$0.5$\pm$0.7 &  4.80$\pm$0.90 & 11.35$\pm$1.94 \\ \hline
\EtaTriPiA&  27 &  2 & 0.2513 &  243 & 35 & 201.7& 0.0639 & 10.3$\pm$0.8$\pm$1.4 &        -       &        -       \\
\EtaTriPiB&  20 &  9 & 0.1820 &   53 &  1 &  50.0& 0.0199 &  8.1$\pm$1.4$\pm$1.6 &        -       &        -       \\
\EtaTriPi &     &    &        &      &    &      &        &  9.5$\pm$0.7$\pm$1.5 &        -       &        -       \\
\EtapTriPi&   1 &  0 & 0.1721 &   17 &  4 &  12.8& 0.0092 &  4.5$\pm$1.6$\pm$1.3 &        -       &        -       \\ \hline
\KKPiPi   & 871 & 83 & 0.2688 & 1072 & 43 & 817.2& 0.3742 &  7.1$\pm$0.3$\pm$0.4 & 16.00$\pm$4.00 &  9.85$\pm$3.23 \\
\RhoKK    & 170 &  - & 0.2602 &  268 &  - & 223.8& 0.3361 &  2.2$\pm$0.2$\pm$0.4 &        -       &        -       \\
\PhiPiPi  &  33 & 13 & 0.2703 &   73 & 20 &  47.6& 0.1744 &  0.9$\pm$0.2$\pm$0.1 &  1.50$\pm$0.28 & 11.07$\pm$3.30 \\
\KKTriPi  & 634 & 18 & 0.2556 &  888 & 19 & 711.6& 0.1818 & 12.7$\pm$0.5$\pm$1.0 &        -       & 10.59$\pm$2.81 \\
\EtaKK    &   3 &  0 & 0.2396 &    7 &  2 &   4.3& 0.0354 &     $ < 1.3 $        &        -       &        -       \\
\OmegaKK  &  62 & 12 & 0.2435 &   97 &  8 &  76.8& 0.1288 &  1.9$\pm$0.3$\pm$0.3 &  1.50$\pm$0.40 & 10.19$\pm$2.96 \\ \hline
\KKKK     & 100 & 11 & 0.2669 &   85 &  2 &  59.2& 0.3118 &  0.6$\pm$0.1$\pm$0.1 &        -       &  6.71$\pm$2.74 \\
\PhiKK    &  46 & 15 & 0.2642 &   49 &  4 &  36.8& 0.1511 &  0.8$\pm$0.2$\pm$0.1 &  0.60$\pm$0.22 &  5.14$\pm$1.53 \\
\KKKKPi   &  20 &  0 & 0.2675 &   51 &  1 &  44.7& 0.1339 &  1.1$\pm$0.2$\pm$0.2 &        -       &        -       \\ \hline
\PPPiPi   & 337 & 28 & 0.2509 & 1010 & 28 & 904.5& 0.4943 &  5.9$\pm$0.2$\pm$0.4 &  8.00$\pm$2.00 &  9.90$\pm$1.16 \\
\RhoPP    &  23 &  - & 0.2570 &   67 &  - &  61.1& 0.4119 &  0.5$\pm$0.1$\pm$0.2 &        -       &        -       \\
\PPTriPi  & 204 &  9 & 0.2312 &  499 & 19 & 434.9& 0.1921 &  7.3$\pm$0.4$\pm$0.6 &        -       & 18.70$\pm$5.80 \\
\EtaPP    &   2 &  1 & 0.2350 &   12 &  2 &   9.8& 0.0399 &  0.8$\pm$0.3$\pm$0.3 &        -       &  3.80$\pm$2.09 \\
\OmegaPP  &  26 &  4 & 0.2173 &   37 & 11 &  21.2& 0.1129 &  0.6$\pm$0.2$\pm$0.2 &  0.80$\pm$0.32 &  4.69$\pm$2.22 \\ \hline
\PPKK     &  25 &  1 & 0.2478 &   37 &  1 &  30.1& 0.3671 &  0.3$\pm$0.1$\pm$0.0 &        -       &        -       \\
\PhiPP    &   2 &  3 & 0.2631 &    6 &  2 &   4.3& 0.1732 &     $ <0.24 $        &   $ <0.26 $    &        -       \\ \hline
\LLPiPi   &  23 &  4 & 0.1902 &   91 & 14 &  73.4& 0.0844 &  2.8$\pm$0.4$\pm$0.5 &        -       &        -       \\
\LamPK    &  65 &  7 & 0.2586 &   97 &  8 &  74.0& 0.2472 &  1.0$\pm$0.1$\pm$0.1 &        -       & 10.92$\pm$2.93 \\
\LamPKPiPi&  29 &  3 & 0.1631 &   57 &  7 &  45.8& 0.0847 &  1.8$\pm$0.3$\pm$0.3 &        -       &        -       \\ \hline
\end{tabular}
\label{tab:br}
\end{center}
\end{table*}

We correct $N_S$ by the efficiency and normalize to the number of
$\psi(2S)$ decays in the data: $3.08\times 10^6$ 
determined by the method described in
\cite{cbx04-35}. The resulting branching ratios are listed in
Table \ref{tab:br}, along with a comparison to the PDG \cite{PDG}.
With the exception of \FourPi\ and \RhoPiPi, none of the 
branching ratios in \cite{PDG} were corrected for the contribution 
from continuum production.

The systematic errors on the ratio of branching fractions share
common contributions from the number of $\psi(2S)$ decays (3.0\%),
uncertainty in $f$ (2.0\%), trigger efficiency (1\%), and Monte
Carlo statistics (2\%). Other sources vary by channel. We include
the following contributions for detector performance modeling
quality: charged particle tracking (0.7\% per track), $\pi^0/\eta
(\to\gamma\gamma)$ finding (4.4\%), $\Lambda$ finding (3\%), $\pi / K/ p$
identification $(0.3\%/1.3\%/1.3\% {\rm ~per~identified~} \pi / K
/p)$, and scaled energy and mass resolutions (2\%). The systematic
error associated with the uncertainty in the level of background
is obtained by recomputing the branching ratio when the background
at the $\psi(2S)$ and the continuum are coherently increased by
$1\sigma$. Since the background in many modes is small, the
Poisson probability for the observed number of background events
to fluctuate up to the 68\% C.L. value is calculated and
interpreted as the uncertainty in the level of background. Many of
the modes studied have resonant submodes, for example: $2(\pi^+
\pi^-)$ where $\rho \pi \pi$ is dominant, $\omega \pi \pi$ where
$b_1 \pi$ is dominant and $\omega f_2(1270)$ is significant,
$K^+K^-\pi^+\pi^-$ where $K^*(892) K \pi$ and $\rho K^+ K^-$ are
large and  $\phi \pi^+ \pi^-$ is small, and $\eta 3\pi$ where
$\eta \pi^0 \rho^0$, $\eta \rho^+ \pi^-$, and $\eta \rho^- \pi^+$
dominate. The efficiencies for the modes with resonant submodes
were weighted using the fractions of the submodes. Allowing for
the presence of resonant submodes changes the efficiency by less
than 5\% relative to the non-resonant efficiency for most modes we
have studied. The difference between the weighted efficiency and
phase space efficiency is taken as a measure of the uncertainty in
the efficiency. For those modes with no observed sub-modes, a 
conservative 10\% uncertainty is assigned for decay model dependence.
Systematic uncertainties are significant for all modes and the
dominant error for many. The measurements in this Letter are in
reasonable agreement with previous measurements where such exist \cite{PDG}.

For the 14 modes where the same final state has been
previously measured at the $J/ \psi$, the $Q_h$ value is computed
using the absolute $\psi(2S)$ branching ratios determined in 
this analysis and $J/\psi$ branching ratios from~\cite{PDG} 
and given in Table \ref{tab:br}. For 5 of these modes:
$K^+K^-\pi^+\pi^- \pi^0 ,~2(K^+K^-),~ p \bar p \pi^+ \pi^- \pi^0
,~ \eta p \bar p$ and \LamPK, this is the first
measurement of $Q_h$.
All modes except $p \bar p \pi^+ \pi^- \pi^0$ have $Q_h$ below 12.7\%.
The modes $\phi K^+ K^-$, $2(\pi^+ \pi^-)$, $2(\pi^+ \pi^-)\pi^0$, 
$\eta p \bar p$, and $\omega p \bar p$ are suppressed with
a significance of 4.7, 4.4, 4.1, 4.1, and 3.5 combined statistical
and systematic standard deviations, respectively.



In summary we have presented first branching fraction measurements 
for 13 modes of the $\psi(2S)$: \EtaTriPi, \EtapTriPi, \RhoKK,
\KKTriPi, \KKKK, \KKKKPi, \RhoPP, \PPTriPi, \EtaPP, \PPKK, \LLPiPi,
\LamPK, \LamPKPiPi, and more precise measurements of 8
previously measured modes:  \FourPi, \RhoPiPi, \FivePi,
\OmegaPiPi, \KKPiPi, \OmegaKK, \PhiKK, \PPPiPi. 
We also measure the branching ratios for \PhiPiPi\ and \OmegaPP\ 
and obtain upper limits for \EtaPiPi, \EtaKK and \PhiPP. 
A full set of measurements of $\pi^+\pi^- + X$, 
$K^+K^- + X$ and $p\bar{p} + X$ where $X$ is 
$\rho/\eta/\omega/\phi$ is included in this analysis.
Results are compared, where possible, with the
corresponding $J/\psi$ branching ratios to test the 12\% rule.
Five modes are inconsistent with the rule, but all values of
$Q_h$ are within a factor of 2 of 12.7\%. Since this analysis
covers a wide variety of final states including modes with and
without strange particles and with and without baryons, the
pattern of branching ratios and $Q_h$ values provide tests of
theoretical models and should allow new insight into the
production widths and the final state interactions operative in
the decays of the $\psi(2S)$.

We gratefully acknowledge the effort of the CESR staff
in providing us with excellent luminosity and running conditions.
This work was supported by the National Science Foundation
and the U.S. Department of Energy.



\begin{thebibliography}{99}
\bibitem{note} We do not veto $J/\psi$ from $\psi(2S) \rightarrow J/\psi \pi^0$
since it is a very small contribution as ${\cal B}(\psi(2S)
\rightarrow J/\psi \pi^0)= (9.6 \pm 2.1) \times 10^{-4}$ \cite{PDG}.
\bibitem{eta_range} For the final state $\eta \pi^+ \pi^- \pi^0$
we require $2.95 < m < 3.15$ GeV.
\end{thebibliography}
\end{document}